\documentclass[11pt,a4paper]{article}

\usepackage[hyperref]{acl2021}

\usepackage{pgfplots}

\newcommand{\tabincell}[2]{\begin{tabular}{@{}#1@{}}#2\end{tabular}}

\usepackage[normalem]{ulem}

\usepackage{times}
\usepackage{latexsym}
\usepackage{amssymb}
\usepackage{amsmath} 
\usepackage{amstext}
\usepackage{booktabs}
\usepackage{enumerate}
\usepackage{graphicx}
\usepackage{subfigure}
\usepackage{xspace}
\usepackage{float}
\usepackage{bbm}
\usepackage{bm}
\usepackage{multirow}
\usepackage{booktabs}
\usepackage{color}
\usepackage{framed}
\usepackage{stfloats}
\usepackage{iitem}
\usepackage[T1]{fontenc}
\definecolor{shadecolor}{RGB}{180,180,180}
\usepackage{url}
\newcommand{\paratitle}[1]{\vspace{1.5ex}\noindent\textbf{#1}}
\newcommand{\ie}{\emph{i.e.,}\xspace}
\newcommand{\aka}{\emph{i.e.,}\xspace}
\newcommand{\eg}{\emph{e.g.,}\xspace}

\newcommand{\ignore}[1]{}

\aclfinalcopy 

\title{PAIR: Leveraging Passage-Centric Similarity Relation for \\ Improving Dense Passage Retrieval}

\author{ \textbf{
Ruiyang Ren\textsuperscript{1,3}\thanks{\llap{}\:\:\:Equal contribution.}  \thanks{\llap{}\:\:\:The work was done when Ruiyang Ren was doing internship at Baidu.} ,
Shangwen Lv\textsuperscript{2}\footnotemark[1] , 
Yingqi Qu\textsuperscript{2}, 
Jing Liu\textsuperscript{2}\thanks{\llap{}\:\:\:Corresponding authors. } ,
Wayne Xin Zhao\textsuperscript{3,4}\footnotemark[3]
} \\
\textbf{
Qiaoqiao She\textsuperscript{2},
Hua Wu\textsuperscript{2},
Haifeng Wang\textsuperscript{2} and Ji-Rong Wen\textsuperscript{3,4}
}\\
	\textsuperscript{1}School of Information, Renmin University of China;  
	\textsuperscript{2}Baidu Inc. \\
	\textsuperscript{3}Beijing Key Laboratory of Big Data Management and Analysis Methods\\
	\textsuperscript{4}Gaoling School of Artificial Intelligence, Renmin University of China\\ 
	\{reyon.ren, jrwen\}@ruc.edu.cn, batmanfly@gmail.com\\
	\{lvshangwen, quyingqi, liujing46, sheqiaoqiao, wu\_hua, wanghaifeng\}@baidu.com \\
}

\begin{document}
\maketitle
\begin{abstract}
Recently, dense passage retrieval has become a mainstream approach to finding relevant information in various natural language processing tasks. A number of studies have been devoted to improving the widely adopted dual-encoder architecture. However, most of the previous studies only consider query-centric similarity relation when learning the dual-encoder retriever. In order to capture more comprehensive similarity relations, we propose a novel approach that 
leverages both query-centric and \textbf{PA}ssage-centric s\textbf{I}milarity \textbf{R}elations (called \textbf{PAIR}) for dense passage retrieval. 
To implement our approach, we make three major technical contributions by introducing formal formulations of the two kinds of similarity relations, generating high-quality pseudo labeled data via knowledge distillation, and designing an effective two-stage training procedure that incorporates passage-centric similarity relation constraint. Extensive experiments show that our approach significantly outperforms previous state-of-the-art models on both MSMARCO and Natural Questions datasets\footnote{Our code is available at ~\url{https://github.com/PaddlePaddle/RocketQA}}.

\end{abstract}

\section{Introduction}
\label{section:intro}

With the recent advances of pre-trained language models, dense passage retrieval techniques (representing queries and passages in low-dimensional semantic space)
have significantly outperformed traditional term-based techniques~\cite{REALM,dpr2020}. 
As the key step of finding the relevant information, it has been shown that dense passage retrieval can effectively improve the performance in a variety of tasks, including question answering~\cite{latent2019acl,multihopqa},  information retrieval~\cite{MEBERT,colbert2020sigir}, dialogue~\cite{conversation,henderson2017efficient} and entity linking~\cite{entitylinking,wu2019scalable}.

\begin{figure}[!tb]
    \centering
	\subfigure[]{\label{fig:motivation_qp2}
 		\centering
		\includegraphics[width=0.201\textwidth]{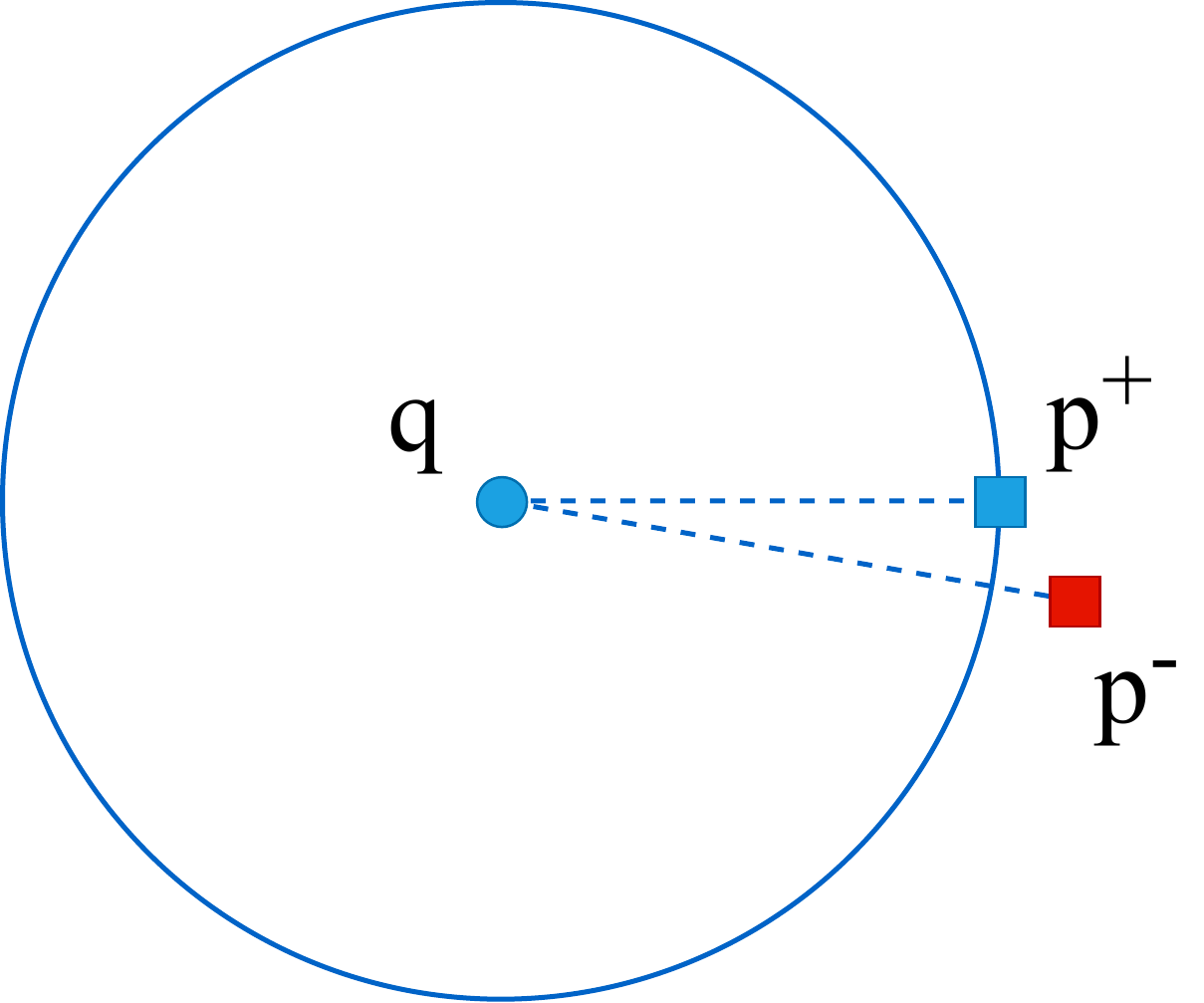}
	}
	\subfigure[]{\label{fig:motivation_pp2}
		\centering
		\includegraphics[width=0.25\textwidth]{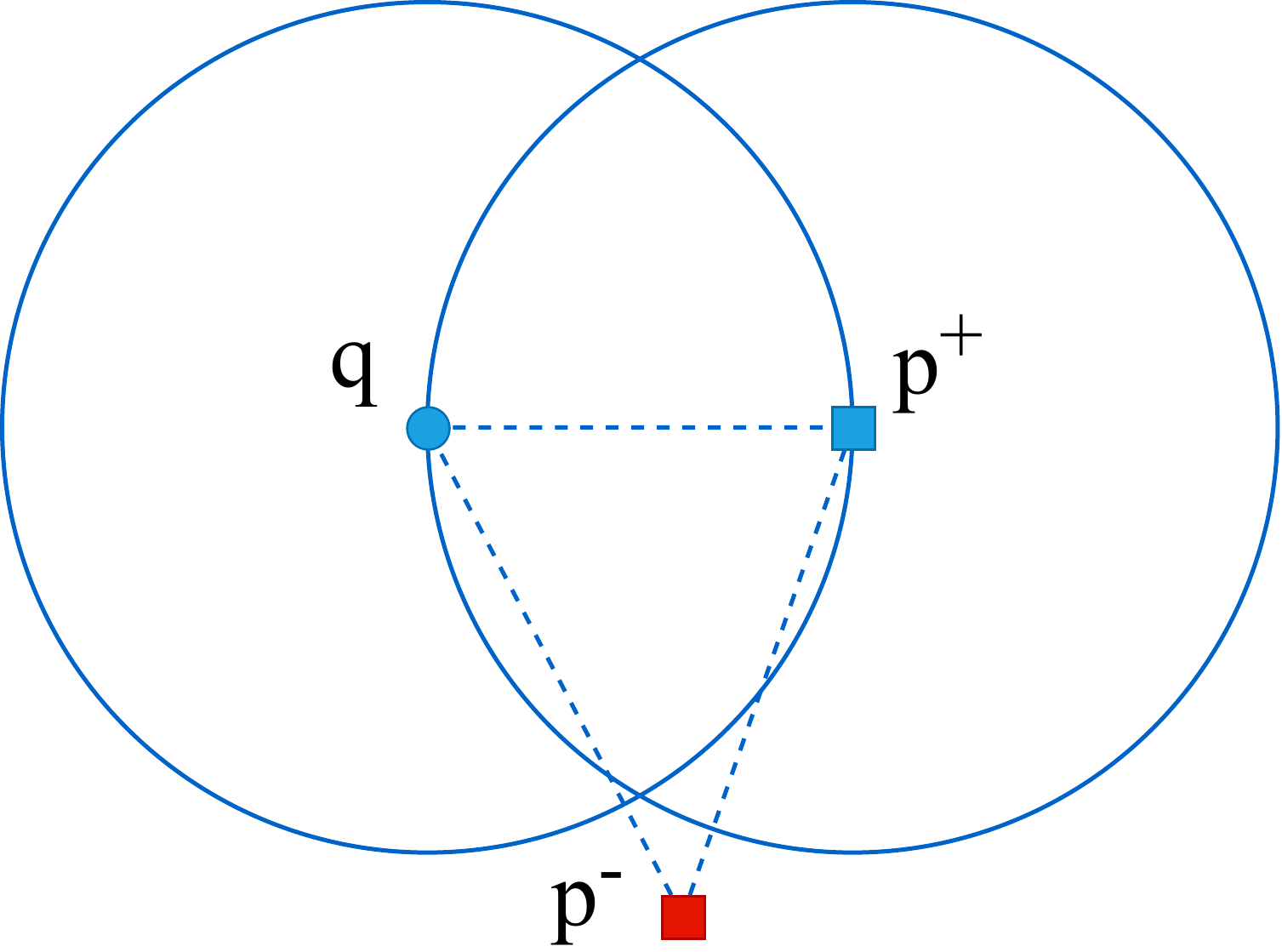}
	}
    \caption{An illustrative case of a query $q$, its positive passage $p^+$ and negative passage $p^-$: (a) Query-centric similarity relation enforces $s(q,p^+) > s(q,p^-)$; (b) Passage-centric similarity relation
    further enforces $s(p^+, q) > s(p^+, p^-)$, where $s(p^+, q) = s(q, p^+)$. We use the distance (\ie dissimilarity) for visualization: the longer the distance is, the less similar it is. 
    }
    \label{fig:intro2}
\end{figure}

Typically, the dual-encoder architecture is used to learn the dense representations of queries and passages, and the dot-product similarity between the representations of queries and passages becomes ranking measurement for retrieval.
A number of studies have been devoted to improving this architecture~\cite{REALM,dpr2020,ANCE} for dense passage retrieval.
Previous studies mainly consider learning query-centric similarity relation, where it tries to increase the similarity $s(q,p^+)$ between a query and a positive (\aka relevant) passage meanwhile decrease the similarity $s(q,p^-)$ between the query and a negative (\aka irrelevant) passage. 
We argue that query-centric similarity relation ignores the relation between passages, and it brings difficulty to discriminate between positive and negative passages. To illustrate this, we present an example in Figure~\ref{fig:intro2}, where a query $q$ and two passages $p^{+}$ and $p^{-}$ are given. As we can see in Figure~\ref{fig:intro2}(a), although query-centric similarity relation can enforce $s(q,p^+) > s(q,p^-)$ and identify the positive passages in this case,  the distance (\ie dissimilarity) between positive and negative passages is small. When a new query is issued, it is difficult to discriminate between positive passage $p^+$ and negative passage $p^-$.  

Considering this problem, we propose to further learn passage-centric similarity relation for enhancing the dual-encoder architecture. The basic idea is shown in Figure~\ref{fig:intro2}(b), where we set an additional similarity relation constraint $s(p^+, q) > s(p^+, p^-)$: the similarity  between query $q$ and positive passage $p^+$ should be larger than that between positive passage $p^+$ and negative passage $p^-$. In this way, it is able to better learn the similarity relations among query, positive passages and negative passages. Although the idea is appealing, it is not easy to implement due to three major issues.  First, it is unclear how to formalize and learn both query-centric and passage-centric similarity relations. 
Second, it requires large-scale and high-quality training data to incorporate passage-centric similarity relation. However, it is expensive to manually label data. Additionally, there might be a large number of unlabeled positives even in the existing manually labeled datasets~\cite{rocketqa}, and it is likely to bring false negatives when sampling hard negatives. 
Finally, learning passage-centric similarity relation (an auxiliary task) is not directly related to the query-centric similarity relation (a target task). In terms of multi-task viewpoint, multi-task models often perform worse than their single-task counterparts~\cite{DBLP:conf/eacl/PlankA17,DBLP:journals/corr/abs-1806-08730,DBLP:conf/acl/ClarkLKML19}. Hence, it needs a more elaborate design for the training procedure. 

To this end, in this paper, we propose a novel  approach that 
leverages both query-centric and \textbf{PA}ssage-centric s\textbf{I}milarity \textbf{R}elations (called \textbf{PAIR}) for dense passage retrieval. In order to address the aforementioned issues, we have made three important technical contributions. First, we design formal loss functions to characterize both query-centric and passage-centric similarity relations. Second, we propose to generate pseudo-labeled data via knowledge distillation. Third, we devise a two-stage training procedure that utilizes passage-centric similarity relation during pre-training and then fine-tunes the dual-encoder according to the task goal.  The improvements in the three aspects make it possible to effectively leverage both kinds of similarity relations for improving dense passage retrieval.

\ignore{We argue that previous works only consider the relation between queries and passages, while ignoring the impact between positive passages and negative passages. As a result, the negatives may be closed to the positives. To illustrate this effect, we use the distance between two texts (an opposite of "similarity") denoted by $\text{d} (\cdot,\cdot)$ in Figure 1. As shown in Figure~\ref{fig:intro2}(a), previous works train the dual-encoders to learn query-centric similarity so that $\text{d}(q,p^+) < \text{d}(q,p^-)$, while $p^+$ and $p^-$ may be closed to each other in the semantic space. Hence, when given a new query, it might be difficult to distinguish between the two passages. To address this issue, it is intuitive to incorporate the passage-centric similarity. Specifically, we mean that the distance between the query and the positive passage should be smaller than the distance between the positive passage and the negative passage, i.e. $\text{d}(p^+, q) < \text{d}(p^+, p^-)$. Overall, we should consider both the query-centric similarity and passage-centric similarity. As shown in Figure~\ref{fig:intro2}(b), the dual-encoders should be optimized so that $\text{d}(q,p^+) < \text{d}(q,p^-)$ and $\text{d}(p^+, q) < \text{d}(p^+, p^-)$. In this paper, we propose an approach that leverages both \textbf{Q}uery-centric and \textbf{P}assage-centric \textbf{S}imilarity (called \textbf{QPSR}) for dense passage retrieval. To implement QPSR, we need to address the following remaining issues and challenges. 
}

The contributions of this paper can be summarized as follows:
\begin{itemize}
    \item We propose an approach that simultaneously learns query-centric and passage-centric similarity relations for dense passage retrieval. It is the first time that passage-centric similarity relation has been considered for this task.
    \item We make three major technical contributions by introducing formal formulations, generating high-quality pseudo-labeled data and designing an effective training procedure. 
    \item Extensive experiments show that our approach significantly outperforms previous state-of-the-art models on both MSMARCO and Natural Questions datasets.
\end{itemize}
\section{Related Work}
Recently, dense passage retrieval has demonstrated better performance than traditional sparse retrieval methods (\eg TF-IDF and BM25). Different from sparse retrieval, dense passage retrieval represents queries and passages into low-dimensional vectors~\cite{zhao2022dense, REALM,dpr2020}, typically in a dual-encoder architecture, and uses dot product as the similarity measurement for retrieval. The existing approaches for dense passage retrieval can be divided into two categories: (1) unsupervised pre-training for retrieval (2) fine-tuning only on labeled data. 

In the first category, different pre-training tasks for retrieval were proposed. \citet{latent2019acl} proposed a specific approach to pre-training the retriever with an unsupervised task, namely Inverse Cloze Task (ICT), and then jointly fine-tuned the retriever and a reader on labeled data. REALM~\cite{REALM} proposed a new pre-training approach, which jointly trained a masked language model and a neural retriever. Different from them, our proposed approach utilizes the pseudo-labeled data via knowledge distillation in the pre-training stage, and the quality of the generated data is high (see Section~\ref{section:sensitivity_analysis}). 

In the second category, the existing approaches fine-tuned pre-trained language models on labeled data~\cite{dpr2020,MEBERT}. 
Both DPR~\cite{dpr2020} and ME-BERT~\cite{MEBERT} used in-batch random sampling and hard negative sampling by BM25, while ANCE~\cite{ANCE}, NPRINC~\cite{google2020negative} and RocketQA~\cite{rocketqa} explored more sophisticated hard negative sampling approach.
~\citet{izacard2020distilling} and ~\citet{google2020augmented} leveraged a reader and a cross-encoder for knowledge distillation on labeled data, respectively. 
RocketQA found large batch size can significantly improve the retrieval performance of dual-encoders. 
ColBERT~\cite{colbert2020sigir} incorporated light-weight attention-based re-ranking while increasing the space complexity. 

The existing studies mainly focus on learning the similarity relation between the queries and the passages, while ignoring the relation among passages. It makes the model difficult to discriminate the positive passages and negative passages. In this paper, we propose an approach simultaneously learn query-centric and passage-centric similarity relations.

\section{Methodology}
In this section, we present an approach that leverages both query-centric and \textbf{PA}ssage-centric s\textbf{I}milarity \textbf{R}elations (called \textbf{PAIR}) for dense passage retrieval.

\subsection{Overview}

The task of dense passage retrieval~\cite{dpr2020} is described as follows. 
Given a query $q$, we aim to retrieve $k$ most relevant passages $\{p_{j}\}_{j=1}^k$ from  a large collection of $M$ passages. 

For this task, the dual-encoder architecture is widely adopted~\cite{dpr2020,rocketqa}, where two separate encoders $E_Q(\cdot)$ and $E_P(\cdot)$ are used to represent the query $q$ and the passage $p$ into $d$-dimensional vectors in different representation spaces. Then a dot product is performed to measure the similarity between $q$ and $p$ based on their embeddings: 
\begin{equation}
\label{equation:sim_definition}
s(q, p)=E_Q(q)^{\top} \cdot E_P(p).
\end{equation}
Previous studies mainly capture the query-centric similarity relation.  
As shown in Figure~\ref{fig:intro2}, passage-centric similarity relation reflects important evidence for improving the retrieval performance. Therefore, we extend the original query-centric learning framework by leveraging the passage-centric similarity relation.

To develop our approach, we need to address the issues described in Section~\ref{section:intro}, and we consider three aspects to extend. First, we design a new loss function that considers both query-centric and passage-centric similarity relations.  Second, we utilize knowledge distillation to obtain large-scale and high-quality pseudo-labeled data to capture more comprehensive similarity relations.  Third, we design a two-stage training procedure to effectively learn the passage-centric similarity relation and improve the final retrieval performance.  

\subsection{Defining the Loss Functions}

\begin{figure}[tbp]
    \centering
    \includegraphics[width=0.48\textwidth]{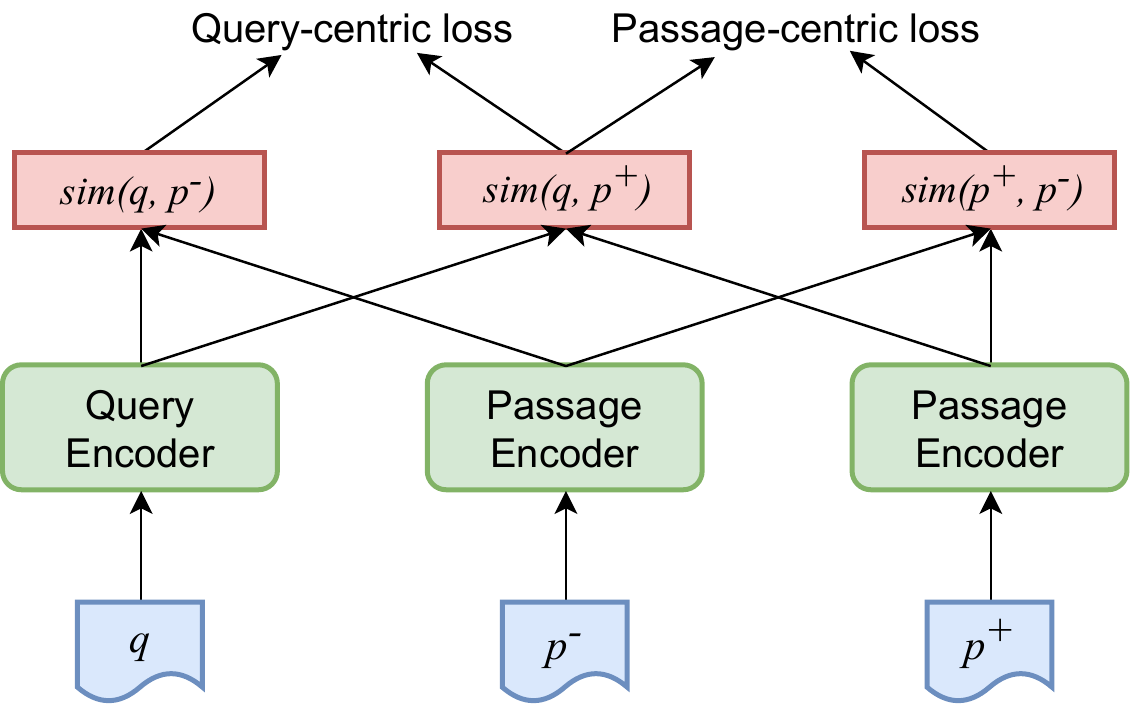}
    \caption{An illustration of the combination of query-centric loss and passage-centric loss.}
    \label{fig:model}
\end{figure}

Our  approach considers two kinds of losses, namely query-centric loss and passage-centric loss, as shown in Figure~\ref{fig:model}.
The two kinds of losses are characterized  by the two different similarity relations, query-centric similarity relation and passage-centric similarity relation.  

\paratitle{Query-centric Loss}\quad The query-centric similarity relation regards the query $q$ as the center and pushes the negative passages $p^-$ farther than the positive passages $p^+$.
That is:
\begin{equation} \label{eq:qp}
    s^{(Q)}(q,p^+) > s^{(Q)}(q,p^-) \,,
\end{equation}
where $s^{(Q)}(q,p^+)$ and $s^{(Q)}(q,p^-)$ represent the similarities for the relevant and irrelevant passages to query $q$, and they are defined the same as $s(q,p)$ in Eq.~\eqref{equation:sim_definition}. 
Following~\cite{dpr2020,rocketqa}, we learn the query-centric similarity relation by optimizing query-centric loss that is the negative log likelihood of the positive passage:
\begin{equation}
\label{equation:L_Q}
\small
L_Q = -\frac{1}{N} \sum_{\langle q, p+\rangle} \log \frac{e^{s^{(Q)}(q, p^+)}}{e^{s^{(Q)}(q, p^+)} + \underline{\sum_{p^-} e^{s^{(Q)}(q, p^-)}}}.
\end{equation}
As shown in Figure~\ref{fig:intro2}, for a given query, there might exist some negative passages similar to the positive passage, making it difficult to discriminate between positive and negative passages. Hence, we further incorporate passage-centric loss to address this issue. 

\paratitle{Passage-centric Loss}\quad The aim of learning passage-centric similarity relation is
to push negative passage $p^-$ farther from positive passage $p^+$, and making the similarity between positive passage $p^+$ and query $q$ larger than the similarity between positive passage $p^+$ and negative passage $p^-$. 
Formally, we introduce the following passage-centric similarity relation: 
\begin{equation} \label{eq:pp}
    s^{(P)}(p^+,q) > s^{(P)}(p^+, p^-),
\end{equation}
where $ s^{(P)}(p^+,q)$ and $s^{(P)}(p^+, p^-)$ are defined as $E_P(p^+)^{\top} \cdot E_Q(q)$ and $E_P(p^+)^{\top} \cdot E_P(p^-)$, respectively. 
Similarly, we learn the passage-centric similarity relation by optimizing the passage-centric loss function that is the negative log likelihood of the query:
\begin{equation}
\label{equation:L_P}
\small
L_P = -\frac{1}{N} \sum_{\langle q, p+\rangle} \log \frac{e^{s^{(P)}(p^+,q)}}{e^{s^{(P)}(p^+,q)} + \underline{\sum_{p^-} e^{s^{(P)}(p^+, p^-)}}}.
\end{equation}
By comparing Eq.~\eqref{equation:L_Q} and Eq.~\eqref{equation:L_P}, we can observe that the difference in two kinds of loss lies in the normalization part (underlined). 

\paratitle{The Combined Loss}\quad
We present an illustrative sketch of the above two loss functions in Figure~\ref{fig:model}. Next, we propose to simultaneously learn both query-centric and passage-centric similarity relations in Eq.(\ref{eq:qp}) and Eq.(\ref{eq:pp}). 
Therefore, we combine query-centric and passage-centric loss functions defined in Eq.~\eqref{equation:L_Q} and \eqref{equation:L_P} to obtain the final loss function:
\begin{equation}
\label{equation:total_loss}
L = (1 - \alpha) * L_Q + \alpha * L_P,
\end{equation}
where $\alpha$ is a hyper-parameter and is tuned in experiments.
By considering passage-centric similarity relation, our approach will be more capable of discriminating between a positive passage and a highly similar yet irrelevant passage \big(See Figure~\ref{fig:intro2}(b)\big). 

\paratitle{Dual-encoder with Shared Parameters}\quad Most of the existing studies~(Eq. (\ref{eq:qp})) equip the dual-encoders with two separate encoders~($E_Q$ and $E_P$) for queries and passages, respectively. In this case, different encoders may project queries and passages into two different spaces. However, to simultaneously model the query-centric similarity relation and the passage-centric similarity relation, the representations of queries and passages should be in the same space. Otherwise, the similarity between passages and the similarity between queries and passages are not comparable. Therefore, we propose using the encoders that share the same parameters and structures for both queries and passages, \ie $E_Q(\cdot)$=$E_P(\cdot)$.

\subsection{Generating the Pseudo-labeled Training Data via Knowledge Distillation}
\label{method_knowledgedistillation}
By optimizing both query-centric loss and passage-centric loss, we can capture more comprehensive similarity relations. However, more similarity relation constraints require large-scale and high-quality training data for optimization. Additionally, there might be a large number of unlabeled positives even in the existing manually labeled datasets~\cite{rocketqa}, and it is likely to bring false negatives when sampling hard negatives. Hence, we propose to generate pseudo-labeled training data via knowledge distillation.  

\paratitle{Cross-encoder Teacher Model}\quad
The teacher model is used to generate large-scale pseudo-labeled data. Following RocketQA~\cite{rocketqa}, we adopt the cross-encoder architecture to implement the teacher, which takes as input the concatenation of query and passage and models the semantic interaction between query and passage representations.  
Such an architecture has been demonstrated to be more effective than the dual-encoder architecture in characterizing query-passage relevance~\cite{google2020augmented}. We follow ~\citet{rocketqa} to train the cross-encoder teacher with the labeled data. 

\paratitle{Generating Pseudo Labels} In this paper, we follow ~\citet{rocketqa} to obtain positives and hard negatives\footnote{\citet{ANCE} and ~\citet{dpr2020} demonstrate the importance of hard negatives.} for unlabeled queries\footnote{We obtain easy negatives from in-batch sampling.}. First, we retrieve the top-\textit{k} candidate passages of unlabeled queries from the corpus by an efficient retriever DPR~\cite{dpr2020}, and score them by the well-trained cross-encoder (\ie teacher model). We set two values $s_{pos}$ and $s_{neg}$ ($s_{pos}>s_{neg}$) as the positive and hard negative thresholds, respectively. Then, given each query, a candidate passage with a score above $s_{pos}$ or below $s_{neg}$ will be considered as positive or negative. 
Note that we also apply this on labeled corpus to obtain more positives and reliable hard negatives. Because there might be a large number of unlabeled positives even in the existing manually labeled datasets~\cite{rocketqa} and it is likely to bring false negatives in hard negative sampling.

\subsection{Two-stage Training Procedure}
\begin{figure}[tbp]
    \centering
    \includegraphics[width=0.48\textwidth]{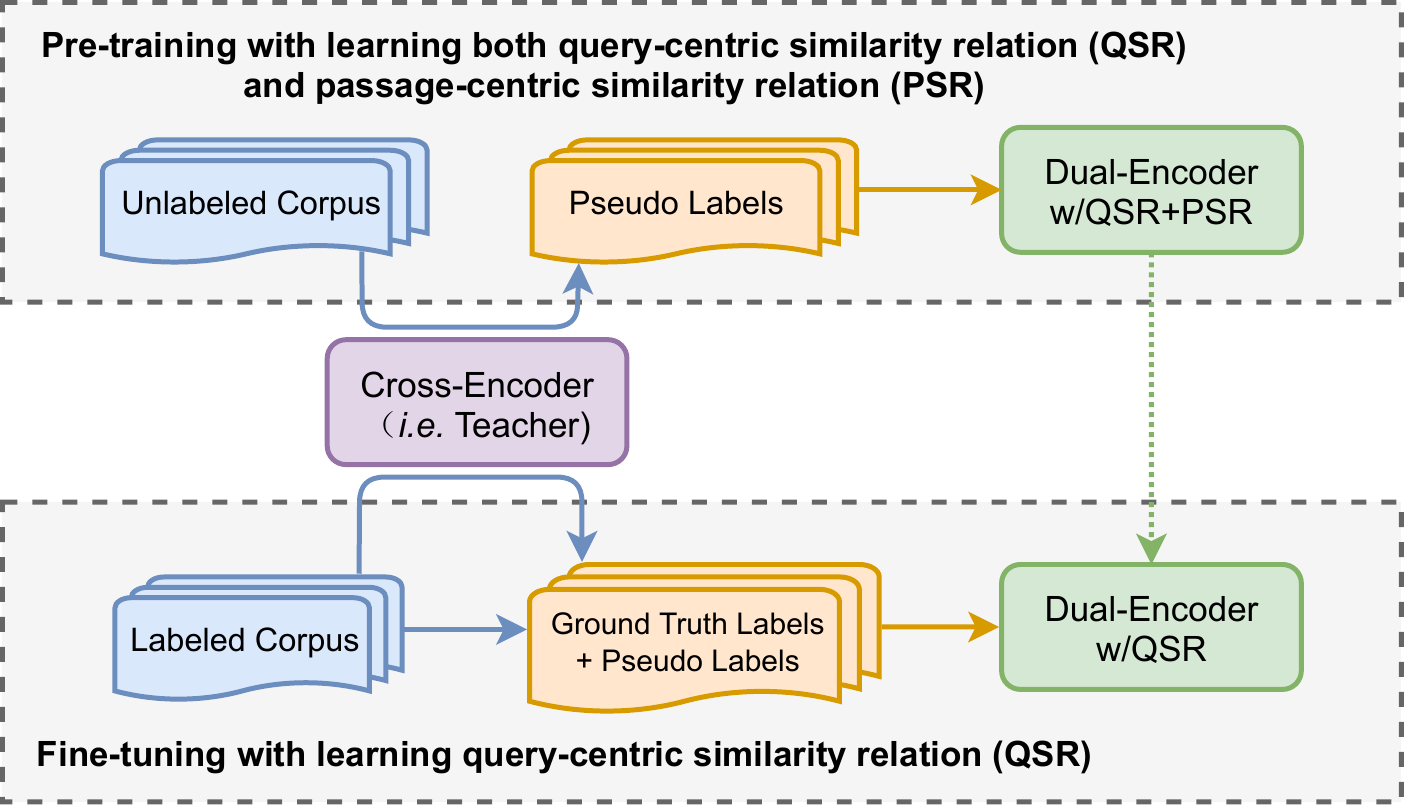}
    \caption{Overview of the proposed two-stage method.}
    \label{fig:pipeline}
\end{figure}

Although passage-centric similarity relation \big(Eq.~\eqref{equation:L_P}\big) is able to incorporate additional relevance evidence, it is not directly related to the final task goal (\ie query-centric similarity relation). Therefore, we design a two-stage training procedure that incorporates the passage-centric loss in the pre-training stage, and then only optimize the tasks-specific loss (\ie query-centric loss) in the fine-tuning stage. We present an illustration for the two-stage training procedure in Figure~\ref{fig:pipeline}. Next, we present the detailed training procedure. 

\paratitle{Pre-training}\quad In the pre-training stage, we train the dual-encoder by optimizing the loss function $L$ in Eq.~\eqref{equation:total_loss} (\ie a combination of query-centric loss and passage-centric loss). The pseudo-labeled data from unlabeled corpus is adopted as the pre-training data~(Section~\ref{method_knowledgedistillation}).

\paratitle{Fine-tuning}\quad In the fine-tuning stage, we only fine-tune the dual-encoder (pre-trained in the first stage) according to the query-centric loss $L_Q$ in Eq.~\eqref{equation:L_Q}. In this way, our approach focuses on learning the task-specific loss, yielding better retrieval performance.  In this stage, we use both ground-truth labels and pseudo labels derived from the labeled corpus for training. 
\begin{table*}
    \centering
    \begin{tabular}{c|c|c|c|c}
    \toprule
     \textbf{Dataset}   &  \textbf{\#q in train} & \textbf{\#q in dev} & \textbf{\#q in test} & \textbf{\#p} \\
     \midrule
        MSMARCO & 502,939 &6,980 & 6.837 & 8,841,823 \\
        Natural Questions & 58,812 & 6,515 & 3,610 & 21,015,324 \\
    \bottomrule
    \end{tabular}
    \caption{The detailed statistics of MSMARCO and Natural Questions. Here, ``q'' and ``p'' are the abbreviations of queries and passages, respectively.}
    \label{tab:dataset}
\end{table*}

\section{Experiments}
In this section, we first describe the experimental settings, then report the main experimental results, ablation study and detailed analysis.

\subsection{Experimental Settings}

\paratitle{Datasets}\quad This paper focuses on the passage retrieval task. We conduct experiments on two public datasets: MSMARCO~\cite{msmarco} and Natural Questions~\cite{nq}. The statistics of the datasets are listed in Table~\ref{tab:dataset}. \emph{MSMARCO} was originally designed for multiple passage machine reading comprehension, and its queries were sampled from Bing search logs. Based on the queries and passages in MSMARCO Question Answering, a dataset for passage retrieval and ranking was created, namely MSMARCO Passage Ranking. \emph{Natural Questions (NQ)} was originally introduced as a dataset for open-domain QA. The queries were collected from Google search logs. DPR~\cite{dpr2020} selected the queries that had short answers, and processed all the Wikipedia articles as the collection of passages. In our experiments, we reuse the version of NQ created by DPR.

\paratitle{Evaluation Metrics}\quad Following previous work, we adopt Mean Reciprocal Rank (MRR) and Recall at top $k$ ranks (Recall@$k$) to evaluate the performance of passage retrieval. MRR calculates the averaged reciprocal of the rank at which the first positive passage is retrieved. Recall@$k$ calculates the proportion of questions to which the top $k$ retrieved passages contain positives.

\paratitle{Unlabeled Corpus}\quad To obtain the augmentation data, we collect about 1.8 million unlabeled queries from Yahoo!~Answers\footnote{\url{http://answers.yahoo.com/}}, ORCAS~\cite{craswell2020orcas}, SQuAD~\cite{rajpurkar2016squad}, TriviaQA~\cite{joshi2017triviaqa} and HotpotQA~\cite{yang2018hotpotqa}. In the pre-training stage, we reuse the passage collections from the labeled corpus (MSMARCO and NQ). 

\begin{table*}[htbp]
    \small
    \centering
    \begin{tabular}{llcccccc}
    \toprule
        \multirow{2}*{\textbf{Methods}} & \multirow{2}*{\textbf{PLM}} &  \multicolumn{3}{c}{\textbf{MSMARCO Dev}} & \multicolumn{3}{c}{\textbf{Natural Questions Test}} \\
               & & MRR@10 & R@50 & R@1000 & R@5 &  R@20 &  R@100\\
    \midrule
        BM25 (anserini)~\cite{bm25} & - & 18.7 & 59.2 & 85.7 & - & 59.1 & 73.7 \\
        \midrule
        doc2query~\cite{doc2query} & - & 21.5 & 64.4 & 89.1 & - & - & - \\
        DeepCT~\cite{deepct2019sigir} & - & 24.3 & 69.0 & 91.0 & - & - & - \\
        docTTTTTquery~\cite{doctttttquery} & - & 27.7 & 75.6 & 94.7 & - & - & - \\
        GAR~\cite{generation20augmented} & - & - & - & - & - & 74.4 & 85.3 \\
        \midrule
        DPR (single)~\cite{dpr2020} & BERT$_\text{base}$ & - & - & - & - & 78.4 & 85.4 \\
        DPR-E & ERNIE$_\text{base}$ & 32.5 & 82.2 & 97.3 &  68.4  & 80.7  & 87.3\\
        ANCE (single)~\cite{ANCE} & RoBERTa$_\text{base}$ & 33.0 & - & 95.9 & - & 81.9 & 87.5 \\
        ME-BERT~\cite{MEBERT} & BERT$_\text{large}$ & 34.3 & - & - & - & - & - \\
        NPRINC~\cite{google2020negative} & BERT$_\text{base}$ & 31.1 & - & 97.7 & 73.3 & 82.8 & 88.4 \\
        ColBERT~\cite{colbert2020sigir} & BERT$_\text{base}$ & 36.0 & 82.9 & 96.8 & - & - & - \\
        RocketQA~\cite{rocketqa} & ERNIE$_\text{base}$ & 37.0 & 85.5 & 97.9 & 74.0 & 82.7 & 88.5 \\
        \midrule
        \textbf{PAIR (Ours)} & ERNIE$_\text{base}$ & \textbf{37.9} & \textbf{86.4} & \textbf{98.2} & \textbf{74.9}  & \textbf{83.5} & \textbf{89.1} \\
    \bottomrule
    \end{tabular}
    \caption{Experimental results on MSMARCO and Natural Questions datasets. Note that we copy the results from original papers and we leave it blank if the original paper does not report the result.}
    \label{tab:main_results}
\end{table*}

\subsection{Implementation Details}
We conduct experiments with the deep learning framework PaddlePaddle~\citep{ma2019paddlepaddle} on up to eight NVIDIA Tesla V100 GPUs (with 32G RAM). 

\paratitle{Pre-trained LMs}\quad The dual-encoder is initialized with the parameters of ERNIE-2.0 base~\citep{ernie20aaai}. ERNIE-2.0 has the same networks as BERT~\cite{bert2019naacl}, and it introduces a continual pre-training framework on multiple pre-trained tasks. The cross-encoder setting follows the cross-encoder in RocketQA~\citep{rocketqa}

\paratitle{Hyper-parameters}\quad (a) \textit{batch size}: Our dual-encoder is trained with a batch size of $512 \times 1$ in fine-tuning stage on NQ and $512 \times 8$ in other settings. We use the in-batch negative setting~\cite{dpr2020} on NQ and cross-batch negative setting~\citep{rocketqa} on MSMARCO. (b) \textit{training epochs}: The number of training epochs is set up to 10 for both pre-training and fine-tuning for dual-encoder. (c) \textit{warm-up and learning rate}: The learning rate of the dual-encoder is set to 3e-5 and the rate of linear scheduling warm-up is set to 0.1. (d) \textit{\# of positives and hard negatives}: The ratio of the positive to the hard negative is set to 1:4 on dual-encoder. 

\paratitle{Optimizers}\quad We use LAMB optimizer~\citep{you2019large} to train the dual-encoder on MSMARCO, which is more suitable in cross-batch negative setting. In other settings, we always use ADAM optimizer~\cite{adam}. 

\paratitle{The choice of alpha}\quad 
$\alpha$ is a hyper-parameter to balance the query-centric loss and passage-centric loss~(Eq.~\eqref{equation:total_loss}). We searched for $\alpha$ from 0 to 1 by setting an equal interval to 0.1, and the model achieves the best performance when $\alpha$ is set to 0.1. 
\subsection{Main Experimental Results}
We consider both sparse and dense passage retrievers for baselines. The sparse retrievers include the traditional retriever BM25~\cite{bm25}, and four traditional retrievers enhanced by neural networks, including doc2query~\cite{doc2query}, DeepCT~\cite{deepct2019sigir}, docTTTTTquery~\cite{doctttttquery} and GAR~\cite{generation20augmented}. Both doc2query and docTTTTTquery employ neural query generation to expand documents. In contrast, GAR employs neural generation models to expand queries. Different from them, DeepCT utilizes BERT to learn the term weight. The dense passage retrievers include DPR~\cite{dpr2020}, DPR-E, ANCE~\cite{ANCE}, ME-BERT~\cite{MEBERT}, NPRINC~\cite{google2020negative}, ColBERT~\cite{colbert2020sigir} and RocketQA~\cite{rocketqa}. DPR-E is our implementation of DPR using ERNIE~\cite{ernie20aaai} instead of BERT, to examine the effects of pre-trained LMs. 

Table~\ref{tab:main_results} presents the main experimental results. 

(1) We can see that PAIR significantly outperforms all the baselines on both MSMARCO and NQ datasets. The major difference between our approach and baselines lies in that we incorporate both query-centric and passage-centric similarity relations, which can capture more comprehensive semantic relations. Meanwhile, we incorporate the augmented data via knowledge distillation. 

(2) We notice that baseline methods use different pre-trained LMs, as shown in the second column of Table~\ref{tab:main_results}. In PAIR, we use the ERNIE-base. To examine the effects of ERNIE-base, we implement DPR-E by replacing BERT-base used in DPR as ERNIE-base. From Table~\ref{tab:main_results}, we can observe that PAIR significantly outperforms DPR-E, although they employ the same pre-trained LM. 

(3) Another observation is that the dense retrievers are overall better than the sparse retrievers. Such a finding has also been reported in prior studies~\cite{dpr2020,ANCE,MEBERT}, which indicates the effectiveness of the dense retrieval approach.

\begin{table}[]
    \fontsize{8.7pt}{0.96em}
    \selectfont
    \centering
    \begin{tabular}{cccc}
    \toprule
    \textbf{Methods} & \textbf{R@5} & \textbf{R@20} & \textbf{R@100} \\
    \midrule
    \tabincell{c}{Complete (PAIR)} & \textbf{74.9}  & \textbf{83.5} & \textbf{89.1} \\
    \midrule
    \tabincell{c}{w/o PSR} & 73.6 &	83.3 & 88.8 \\
    \tabincell{c}{w/o KD} & 70.9 & 82.7 & 88.1 \\
    \tabincell{c}{w/ PSR FT} & 74.6 & 83.4 & 89.0  \\
    \tabincell{c}{w/o SP} & 74.0	& 83.4	& 88.9 \\
    \tabincell{c}{w/o PT} & 73.0 &	82.8 &	88.5 \\
    \bottomrule
    \end{tabular}
    \caption{The ablation study and controlled experiments of different variants of PAIR on Natural Questions. }
    \label{tab:ablation_study}
\end{table}

\subsection{Ablation Study}
In this section, we conduct ablation study to examine the effectiveness of each strategy in our proposed approach. We only report the results on the NQ, while the results on the MSMARCO are similar and omitted here due to limited space. 

Here, we consider five variants based on our approach for comparison: 

(a) \underline{\emph{w/o PSR}} removes the loss for passage-centric similarity relation in the pre-training stage; 

(b) \underline{\emph{w/o KD}} removes the knowledge distillation for obtaining pseudo-labeled data and only uses the labeled data (MSMARCO and NQ) for both pre-training stage and fine-tuning stage; 

(c) \underline{\emph{w/ PSR FT}} adds the loss for passage-centric similarity relation in the fine-tuning stage;

(d) \underline{\emph{w/o SP}} uses separate encoders for queries and passages instead of encoders with shared parameters; 

(e) \underline{\emph{w/o PT}} removes the pre-training stage. 

Table~\ref{tab:ablation_study} presents the results for the ablation study.
We can observe the following findings: 

$\bullet$ The performance  drops in \underline{\emph{w/o PSR}}, demonstrating the effectiveness of learning passage-centric similarity relation;

$\bullet$ The performance drops  in \underline{\emph{w/o KD}}, demonstrating the necessity and effectiveness of the knowledge distillation for obtaining large-scale and high-quality pseudo-labeled data, since the passage-centric loss tries to distinguish highly similar but semantically different passages;

$\bullet$ The performance slightly drops in \underline{\emph{w/ PSR FT}}, because  passage-centric loss is not directly related to the target task (\ie query-based retrieval), which suggests that passage-centric loss should be only used in the pre-training stage;

$\bullet$ The performance drops in \underline{\emph{w/o SP}}, demonstrating the effectiveness of dual-encoders with shared parameters;

$\bullet$ The performance significantly drops in \underline{\emph{w/o PT}}, demonstrating the importance of our pre-training procedure. 

\begin{figure}
	\centering
	\includegraphics[width=0.43\textwidth]{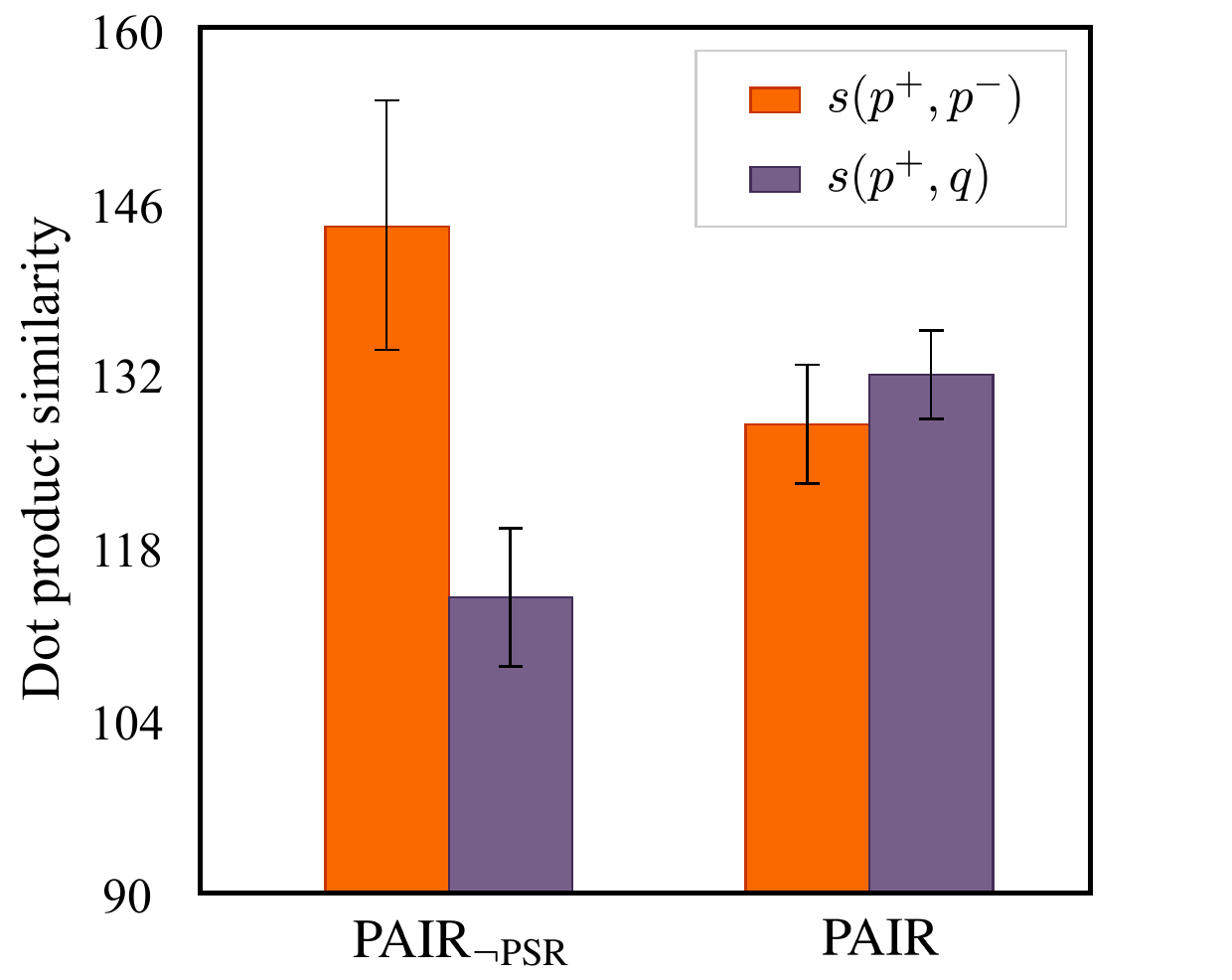}
	\caption{The comparison of PAIR and PAIR$_{\neg \text{PSR}}$ on $s(p^+, p^-)$ and $s(p^+,q)$ with standard deviation.}
	\label{fig:quantitative}
\end{figure}

\begin{table*}[!t]
\scriptsize
\begin{tabular}{p{3cm}|p{6cm}|p{6cm}}

\toprule
Query & Top 1 passage retrieved by PAIR (correct) & Top 1 passage retrieved by PAIR$_{\neg \text{PSR}}$ (incorrect) \\
\midrule
\multirow{1}{3cm}[-0.6em]{Which animal is the carrier of the \textbf{H1N1} virus ?} 
& \underline{\textbf{H1N1} strains caused a small percentage of all human flu} \underline{infections} in 2004–2005. Other strains of \textbf{H1N1} are endemic \uwave{\textit{ in pigs}} (swine influenza) and in birds (avian influenza) \ldots 
 & \underline{\textbf{H5N1} is a subtype virus which can cause illness in humans} and many other animal species. A bird-adapted strain of \textbf{H5N1}, called HPAIA (\textbf{H5N1}) for  \ldots
 \\
 \midrule
\multirow{1}{3cm}[-0.6em]{Where is \textbf{gall bladder} situated in human body?
} 
& \uline{The \textbf{gall bladder} is a small hollow organ} where bile is stored \ldots In humans, the pear-shaped \textbf{gall bladder} lies \uwave{\textit{beneath the liver}}, although the structure and position \ldots
& \uline{The \textbf{urinary bladder} is a hollow muscular organ} in humans and some other animals that collects and stores urine from the kidneys before disposal by urination \ldots \\
\bottomrule
\end{tabular}
\caption{The comparison of the top-1 passages retrieved by PAIR and PAIR$_{\neg \text{PSR}}$, respectively. The \textbf{bold words} represent the main topics in queries and passages. The \uwave{\textit{italic words with wavy underline}} are the right answers. The \uline{words with straight underline} among passages have many words in common and may mislead the model PAIR$_{\neg \text{PSR}}$ to select the wrong passage.}
\label{tab:case-study}
\end{table*}

\subsection{Analysis on Passage-centric Similarity Relation}
The previous results demonstrate the effectiveness of our proposed approach PAIR. Here, we further analyze the effect of passage-centric loss (Eq.~\eqref{equation:L_P}) in a more intuitive way. 
To examine this, we prepare two variants of our approach, namely the complete PAIR and the variant removing the passage-centric loss (Eq.~\eqref{equation:L_P}) denoted by PAIR$_{\neg \text{PSR}}$.

We first analyze how the passage-centric similarity relation (PSR) influences the similarity relations among query, positive passage and negative passage. 
 Figure~\ref{fig:quantitative} shows the comparison of PAIR and PAIR$_{\neg \text{PSR}}$ for computing the similarities of $s(p^+, p^-)$ and $s(p^+,q)$. We obtain $s(p^+, p^-)$ and $s(p^+,q)$ by the averaging the similarity of top 100 retrieved passages for each query in the testing data of Natural Questions. We can see that before incorporating passage-centric similarity relation~(PSR), $s(p^+, p^-)$ is higher than $s(p^+,q)$. 
 As a result, the negatives are close to the positives. After incorporating PSR, $s(p^+, p^-)$ becomes lower than $s(p^+,q)$. It indicates that passage-centric loss pulls positive passages closer to queries and push them farther away from negative passages in the representation space. The comparison result is consistent with passage-level similarity relation in Eq.~(\ref{eq:pp}). 

Next, we further present two examples in Table~\ref{tab:case-study} to understand the performance difference between PAIR and PAIR$_{\neg \text{PSR}}$.
In the first example, the top-1 passage retrieved by PAIR has the same topic ``H1N1'' as the query. In contrast, the top-1 passage retrieved by PAIR$_{\neg \text{PSR}}$ has an incorrect but highly relevant topic ``H5N1''. Actually, the sentences among the positive passage (retrieved by PAIR) and the negative passage (retrieved by PAIR$_{\neg \text{PSR}}$) share many common words. Such a negative passage is likely to mislead the retriever to yield incorrect rankings. Hence, these two passages should be far away from each other in the representation space. This problem cannot be well solved by only considering the query-passage similarity in existing studies.
Similar observations can be find from the second example.
The top-1 passage retrieved by PAIR has the same topic ``gall bladder'' as the query, while the top-1 passage retrieved by PAIR$_{\neg \text{PSR}}$ is about ``urinary bladder''. 
These results show that passage-centric similarity relations are particularly useful to discriminate between positive and hard negative passages (highly similar to positive passages).

\begin{table}
    \fontsize{7.7pt}{0.82em}
    \selectfont
    \centering
    \begin{tabular}{c|cc|ccc}
        \toprule
        \multirow{2}*{\textbf{Threshold}} &  \multicolumn{2}{c|}{\textbf{Data Quality}} & \multicolumn{3}{c}{\textbf{Retrieval Performance}} \\ 
        
        & Acc$_{pos}$ & Acc$_{neg}$ & R@5 & R@20 & R@100 \\
        \midrule
        \tabincell{c}{$s_{pos}=0.9$} & \multirow{2}*{\textbf{92\%}} & \multirow{2}*{\textbf{96\%}}
        & \multirow{2}*{\textbf{74.9}} & \multirow{2}*{\textbf{83.5}} & \multirow{2}*{\textbf{89.1}} \\
        \tabincell{c}{$s_{neg}=0.1$}& & & & &\\
        \midrule
        \tabincell{c}{$s_{pos}=0.8$} & \multirow{2}*{90\%} & \multirow{2}*{93\%} 
        & \multirow{2}*{74.5} & \multirow{2}*{83.4} & \multirow{2}*{88.9} \\ 
        \tabincell{c}{$s_{neg}=0.2$}& & & & &\\
        \midrule
        \tabincell{c}{$s_{pos}=0.7$} & \multirow{2}*{84\%} & \multirow{2}*{87\%}
        & \multirow{2}*{73.6} & \multirow{2}*{\textbf{83.5}} & \multirow{2}*{88.6} \\
        \tabincell{c}{$s_{neg}=0.3$} & & & & &\\
        \midrule
        \tabincell{c}{$s_{pos}=0.6$} & \multirow{2}*{80\%} & \multirow{2}*{87\%}
        & \multirow{2}*{73.5} & \multirow{2}*{83.4} & \multirow{2}*{88.7} \\
        \tabincell{c}{$s_{neg}=0.4$} & & & & &\\
        \bottomrule
    \end{tabular}
    \caption{The data quality and retrieval performance in different thresholds on NQ. Acc$_{pos}$ denotes accuracy of positives and Acc$_{neg}$ denotes accuracy of negatives. }
    \label{tab:distill}
\end{table}

\subsection{Analysis on Knowledge Distillation}
\label{section:sensitivity_analysis}
In this section, we examine the influence of the thresholds on pseudo-labeled data via knowledge distillation, including the data quality and the retrieval performance. We conduct the analyses by using different positive thresholds $s_{pos}$ and negative thresholds $s_{neg}$ (See Section~\ref{method_knowledgedistillation}). 

We first manually evaluate the quality of the pseudo-labeled data via knowledge distillation \emph{w.r.t.} different threshold settings (\ie the combinations of $s_{neg}$ and $s_{pos}$). For each threshold setting, we randomly select 100 queries, each of which corresponding to a positive passage and a hard-negative passage. In total, we have 4 threshold settings (as shown in Table~\ref{tab:distill}) 
and 800 query-passage pairs. We ask two experts to manually annotate the query-passage pairs and evaluate the quality of pseudo-labeled data, the Cohen's Kappa of experts is 0.9. 
As shown in the first two columns of Table~\ref{tab:distill}, we can observe that when $s_{pos} = 0.9$ and $s_{neg} = 0.1$, the data quality is relatively good. Additionally, when setting a low value of $s_{pos}$ and a high value of $s_{neg}$, the data quality becomes worse. 

The last three columns of Table~\ref{tab:distill} also present the retrieval performance \emph{w.r.t}. different threshold settings. When choosing a low value of $s_{pos}$ and a high value of $s_{neg}$, the retrieval performance drops. Hence, our approach is configured with a strict threshold setting~($s_{pos}$ = 0.9, $s_{neg}$ = 0.1) in experiments to achieve good performance.

\section{Conclusion and Future Work}
This paper presented a novel dense passage retrieval approach that  leverages both query-centric and passage-centric similarity relations for capturing more comprehensive semantic relations. 
To implement our approach, we made three important technical contributions in the loss formulation, training data augmentation and effective training procedure. Extensive results demonstrated the effectiveness of our approach. 
To our knowledge, it is the first time that passage-centric similarity relation has been considered for dense passage retrieval. We believe such an idea itself is worth exploring in designing new ranking mechanism. In future work, we will design more principle ranking functions and apply current retrieval approach to downstream tasks such as question answering and passage re-ranking. 

\section{Ethical Impact}
The technique of dense passage retrieval is effective for question answering, where the majority of questions are informational queries. Semantic crowdedness problem of passages, and term mismatch between questions and passages are typical problems, which bring barriers for the machine to accurately find the information.
Our technique contributes toward the goal of asking machines to find the answer passages to natural language questions from a large collection of documents. 
With these advantages also come potential downsides: Wikipedia or any potential external knowledge source will probably never fully cover the breadth of user questions. The goal is still far from being achieved, and more efforts from the community is needed for us to get there.

\section*{Acknowledgement}
This work is partially supported by the National Key Research and Development Project of China (No.2018AAA0101900), National Natural Science Foundation of China under Grant No. 61872369 and Beijing Outstanding Young Scientist Program under Grant No. BJJWZYJH012019100020098.

\bibliography{acl2021.bib}
\bibliographystyle{acl_natbib.bst}
 
\end{document}